\documentclass[aps,prl,preprint]{revtex4-1}

\usepackage{graphicx}
\usepackage{amsmath}
\usepackage{mathtools}
\usepackage{braket}
\usepackage{bbold}
\usepackage{units}

\begin{document}

\title{Observation of enhanced rate coefficients in the H$_2^+$\ +\ H$_2$ $\rightarrow$ H$_3^+$\ +\ H reaction at low collision energies}

\author{Pitt Allmendinger\footnote{The first two authors contributed equally to this work}}
\author{Johannes Deiglmayr\footnotemark[1]}
\author{Katharina H{\"o}veler}
\author{Otto Schullian}
\author{Fr{\'e}d{\'e}ric Merkt\footnote{Corresponding author}}
\affiliation{Laboratory of Physical Chemistry, ETH Zurich, Zurich, Switzerland}
\email{address: merkt@phys.chem.ethz.ch}

\begin{abstract}
The energy dependence of the rate coefficient of the H$_2^+\ + {\rm H}_2 \rightarrow {\rm H}_3^+ + {\rm H}$ reaction has been measured in the range of collision energies between $k_\mathrm{B}\cdot\unit[10]{K}$ and $k_\mathrm{B}\cdot\unit[300]{mK}$. A clear deviation of the rate coefficient from the value expected on the basis of the classical Langevin-capture behavior has been observed at collision energies below $k_\mathrm{B}\cdot\unit[1]{K}$, which is attributed to the joint effects of the ion-quadrupole and Coriolis interactions in collisions involving ortho-H$_2$ molecules in the $j = 1$ rotational level, which make up 75\% of the population of the neutral H$_2$ molecules in the experiments. The experimental results are compared to very recent predictions by Dashevskaya, Litvin, Nikitin and Troe (J. Chem. Phys., accompanying article), with which they are in agreement.
\end{abstract}

\maketitle
\section{introduction}\label{intro}
Ion-molecule reactions represent a specific class of chemical reactions that differ from neutral-neutral reactions in several important aspects \cite{bowers79a,ng92a,clary90a,gerlich92a,anderson01a,smith11a,oka13a}: Firstly, many important ion-molecule reactions are barrierless and exothermic, and are therefore fast, even at low temperatures. Secondly, the attraction forces between a nonpolar molecule and an ion are dominated, at long range, by the ion--induced-dipole interaction. The interaction potential scales with the intermolecular distance $R$ as $R^{-4}$ and extends to much larger distances than the potential between two neutral species. Finally, the centrifugal barriers in the intermolecular potential associated with different collisional partial waves are lowered by the long-range attraction so that at a given collisional energy more partial waves contribute to an ion-neutral collision than to the corresponding neutral-neutral collision. Ion-neutral collisions can thus often be described by classical dynamics at temperatures where quantum mechanical effects dominate the corresponding neutral-neutral collision.

Studies of chemical processes involving ions at low temperature are motivated in part by the need to understand and model chemical reaction cycles in interstellar molecular clouds \cite{herbst73a,gerlich06a,smith11a,oka13a}, which are characterized, depending on the nature of the cloud, by temperatures down to the cosmic background temperature of 2.7~K. They are also motivated by the desire to explore the regime of cold and ultracold chemistry where the reactivity is influenced by quantum phenomena~\cite{krems09a,bell09a,meerakker12a,narevicius12a,heazelwood15a}.

Ion-molecule reactions can often be described, at low temperatures, by capture models (see, e.g., \cite{gioumousis58a,bowers79a,su82a,clary85a,troe87a,clary90a,troe96a,dashevskaya05a,gao11a,dashevskaya16a}). Such models rely on the assumption that the rate coefficients only depend on the long-range electrostatic interactions between the charge $q$ of the ion and the induced (polarizability $\alpha$) or permanent electric dipoles, quadrupoles, etc., of the neutral species and do not consider the details of the chemical transformation, which are determined by short-range interactions. Capture models usually express the rate coefficients $k$ as ratios $\frac{k}{k_{\rm L}}$ to the classical Langevin-capture rate coefficient \cite{langevin05a} (in SI units)
\begin{equation}\label{eq:langevin}
k_{\rm L}=\sqrt{\frac{\alpha q^2}{4\epsilon_0^2 \mu}}=v_{\rm rel}\sigma_{\rm L},
\end{equation}
which is temperature independent and known to provide a good description of exothermic, barrier-free reactions between ions and polarizable molecules (e.g., H$_2$ or N$_2$) down to very low temperatures \cite{mackenzie94a,glenewinkelmeyer97a,sanzsanz15a}, even below 1~K \cite{dashevskaya05a}. In Eq.~\eqref{eq:langevin}, $\alpha$, $q$, $\epsilon_0$, $\mu$, $v_{\rm rel}$, and $\sigma_{\rm L}$ are the polarizability of the neutral reactant, the charge of the ionic reactant, the electric constant, the reduced mass and the relative velocity of the collision partners, and the Langevin cross section, respectively. In the zero-collision-energy limit, the capture rate coefficients must deviate from the classical Langevin behavior and reach their quantum (q) s-wave-scattering values~$k_{{\rm q}}$ \cite{landau77a}. For collisions of ions with neutral atoms or molecules without permanent moments, for instance, $\frac{k_{{\rm q}}}{k_{\rm L}}=2$ \cite{vogt54a,fabrikant01a,dashevskaya05a,gao11a}.

As the collisional temperature rises above 0 K, the number of partial waves contributing to the scattering increases until the relative motion of the reactants can be described within the classical approximation. The transition from quantum ($k/k_{\rm L}=2$) to classical ($k/k_{\rm L}=1$) capture takes place in the sub-Kelvin range, even for ion-molecule reactions involving the lightest species, and has not been observed experimentally so far. For collisions of ions with neutral diatomic molecules, the rotational degrees of freedom of the neutral molecule get gradually excited in the range from 0.1 to 20 K and unlock themselves from, but remain strongly perturbed by, the anisotropic intermolecular potential and by Coriolis interactions \cite{dashevskaya05a,maergoiz09a}. In this range, which also remains unexplored experimentally, strongly enhanced and quantum-state-specific rate coefficients are predicted theoretically \cite{clary85a,troe87a,clary90a,wickham93a,troe96a,auzinsh08a,klippenstein10a,auzinsh13a,auzinsh13b,dashevskaya16a}.

We present here a measurement of the energy dependence of the cross section of the reaction
\begin{equation}\label{eq:reaction}
{\rm H}_2^+ + {\rm H}_2\rightarrow {\rm H}_3^+ + {\rm H}
\end{equation}
in the range of collision energies from $k_\mathrm{B}\cdot\unit[10]{K}$ to $k_\mathrm{B}\cdot\unit[300]{mK}$ ($k_\mathrm{B}$ is Boltzmann's constant). When the collision energy is reduced below $k_\mathrm{B}\cdot\unit[1]{K}$, we observe an increase of the rate coefficient and a clear deviation from the classical capture-rate coefficient. At a collision energy of $k_\mathrm{B}\cdot\unit[1]{K}$, partial waves up to $\ell=5$ contribute to the collision. On the basis of the predictions of Ref.~\cite{dashevskaya05a} and the results presented in the accompanying article \cite{dashevskaya16a}, we attribute the observed enhancement to the joint effects of the ion-quadrupole and Coriolis interactions in collisions involving ortho-H$_2$ molecules in the $j = 1$ rotational level, which make up 75\% of the beam intensity in our experiments.

\section{Experimental procedure}

\begin{figure}
\begin{center}
\includegraphics[width=\columnwidth]{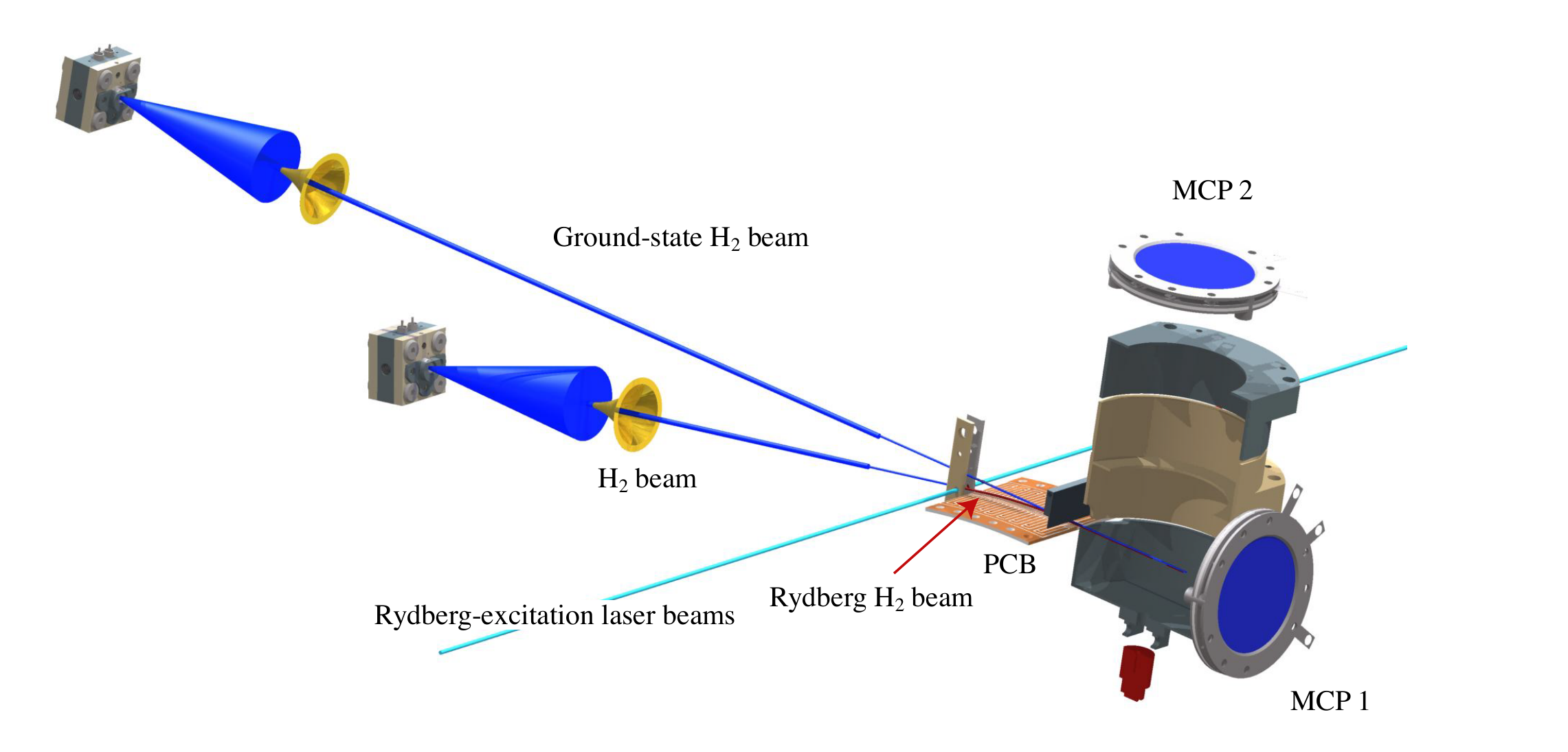}
\caption{\label{fig:setup}Schematic representation of the merged-beam apparatus used to study ion-molecule reactions at low collision energies, with the two skimmed supersonic beams initially propagating at an angle of 10$^\circ$, the Rydberg-Stark deflector made of a bent printed circuit board (PCB) and used to merge the beams after laser excitation, the reaction zone located within an electrode stack (gray), which constitutes the linear time-of-flight spectrometer used to detect reactants and products separately. (MCP1) and (MCP2) Microchannel-plate detectors to monitor the flight times of Rydberg H$_2$ molecules and the ion-time-of-flight spectra, respectively.}
\end{center}
\end{figure}

The experiments are performed with a merged-beam apparatus designed for studies of ion-molecule reactions at low collision energies ~\cite{allmendinger16a}. The experimental setup is depicted schematically in Fig.~\ref{fig:setup}. Two supersonic molecular beams of H$_2$ are created by expansion of H$_2$ from liquid-nitrogen-cooled pulsed valves and initially propagate at an angle of 10$^\circ$. The molecules in one of these two beams are excited from the $\mathrm{X}\, ^1\Sigma_{\mathrm g}^+ \, (v=0,j=0)$ ground state to long-lived $nkm$ Rydberg-Stark states (designated H$_2^*$ below) with principal quantum number $n=22$ and the ion core in the $\mathrm{X}^+\, ^2\Sigma_{\mathrm g}^+\,(v^+=0,N^+=0)$ state using a resonant three-photon excitation sequence~\cite{seiler11b}. The two beams are then merged by deflecting the Rydberg molecules with a surface-electrode Rydberg-Stark deflector and accelerator (RYSDAC)~\cite{allmendinger13a}.

Once the two beams are merged, they co-propagate through the reaction zone (see Fig.~\ref{fig:setup}), where the reaction 
\begin{equation}\label{rydberg-reaction}
{\rm H}_2^* + {\rm H}_2\rightarrow {\rm H}_3^+ + {\rm H} + {\rm e}^-
\end{equation}
is observed by monitoring the yield of H$_3^+$ ions after extraction with a pulsed electric field and detection at a microchannel-plate detector located at the end of a time-of-flight mass-spectrometer (MCP2). The cross sections of Reaction~(\ref{rydberg-reaction}) are equivalent, within the sensitivity limits of our experiment, to the cross section of Reaction~\eqref{eq:reaction}, which can be regarded as taking place within the orbit of the distant Rydberg electron without being affected by it, 
as was demonstrated previously~\cite{pratt94a,matsuzawa10a,allmendinger16a}. This equivalence, which has also been exploited in studies of the H$^+ + {\rm H}_2$ reaction \cite{wrede05a,dai05a}, can be rationalized by comparing the maximal impact parameter $r_L=\sqrt{\sigma_{\rm L}/\pi}$ of the ion-neutral reaction to the classical Rydberg-orbit radius $\langle r \rangle_n =  a_0n^2$. Even at the lowest collision energies ($k_\mathrm{B}\cdot\unit[300]{mK}$) investigated here, $\langle r \rangle_{n=22}$ is more than ten times larger than $r_L$. Moreover, in the long-lived Rydberg molecules that survive the time interval of more than 50 $\mu$s between initial preparation by laser excitation and the reaction, the Rydberg-electron density close to the ion core is negligible. Consequently, the Rydberg electron does not influence the reaction but acts like a Faraday cage, which shields the charged reactants from heating by external stray fields and allows the control of the collision energy with the  precision achievable in neutral-neutral reactions~\cite{henson12a,shagam15a,jankunas14a}. This control is achieved by varying the temperature of the pulsed valve generating the beam of ground-state molecules while adjusting the valve trigger time for optimal overlap of the two merged packets (\textit{i.e.}, for maximal number $N_{\mathrm{H}_3^+}^\mathrm{max}$ of detected H$_3^+$ ions), or by acceleration of the H$_2^*$ molecules using the RYSDAC. 

The distribution of collision energies is determined from the time- and position-dependent velocity distributions of the two beams. The packet of ground-state H$_2$ molecules has dispersed strongly when reaching the reaction zone. This dispersion, which leads to a strong correlation between spatial position and velocity, results from the short opening time of the pulsed valve (13~$\mu$s) and the long distance of about 80~cm between the orifice of the valve and the reaction zone. The packet of Rydberg molecules, in contrast, is strongly localized when released from the RYSDAC and disperses only weakly, corresponding to a translational temperature of approximately 300~mK. The H$_2^*$ sample thus remains strongly localized in space as it propagates through the sample of ground-state H$_2$ molecules in the reaction zone. At very low mean collision energies, the distribution of collision energies, and thus the energy resolution of the measurement, is limited by the translational temperature of the H$_2^*$ Rydberg molecules. For higher relative velocities of the two beams, the energy resolution depends on the time during which the two beams are allowed to interact before the product ions are extracted. To adjust this time, the H$_3^+$ ion yield is measured by using either a one-pulse or a two-pulse extraction-field sequence, the latter permitting a higher collision-energy resolution, as explained in Ref.~\cite{allmendinger16a}.

Under conditions where the opening time of the ground-state H$_2$ valve is minimized, we observe a weak, second gas pulse originating from the rebounce of the valve plunger. This rebounce pulse has been fully characterized  (see Ref.~\cite{allmendinger16a}). For all experimental results presented below, we have either fully included its effects in the analysis or made sure that its effects are negligible.

The peak number density of H$_2$ molecules greatly exceeds that of the H$_2^*$ Rydberg molecules. The reaction probability per Rydberg molecule is, however, much smaller than one. The reaction thus follows pseudo-first-order kinetics and the number of formed H$_3^+$ ions is directly proportional to the rate coefficient $k$, the number $N_{\mathrm{H}_2}$ of H$_2$ molecules and $N_{\mathrm{H}_2^*}$ of H$_2^*$ molecules, and to the product, averaged over time ($t$) and volume ($V$),
of the two normalized density distributions, which represents an overlap integral over the reactant distributions
\begin{equation}\label{eq:Nh3}
N_{\mathrm{H}_3^+} = k \, N_{\mathrm{H}_2} \, N_{\mathrm{H}_2^*} \langle \rho_{\mathrm{H}_2} \rho_{\mathrm{H}_2^*}\rangle_{t,V,\gamma}. \;
\end{equation}
The relative number of H$_2$ molecules as a function of the valve temperature is determined in a separate measurement by electron-impact ionization directly in the reaction zone, while the relative number of H$_2^*$ molecules is obtained from the pulsed-field-ionization signal monitored at MCP2 which is recorded together with the H$_3^+$  ion signal (see Fig.~\ref{fig:setup} and Ref.~\cite{allmendinger16a}).  The overlap factor $\langle \rho_{\mathrm{H}_2} \rho_{\mathrm{H}_2^*}\rangle_{t,V,\gamma}$ depends on the experimental parameters (indicated by the index $\gamma$) and is obtained, together with the mean collision energy and energy resolution, by Monte-Carlo particle-trajectory simulations of the two beams. The simulations are based on the experimentally measured velocity and density distributions of the H$_2$ ground-state beam, the well-defined electric-potential functions applied to the RYSDAC, and the exact geometry and timings of the experiment, as described in Ref.~\cite{allmendinger16a}. Additionally, the efficiency with which a product ion is detected depends on the center-of-mass velocity of the reactants, the velocity and propagation direction of the product ion, and the time interval between formation and extraction of the product ion.
The detection efficiency $D_\gamma$ is also determined by Monte-Carlo simulations of the reaction and detection process~\cite{allmendinger16a}. The relative rate coefficient of Reaction~(\ref{eq:reaction}) is then obtained by dividing the measured number $N_{\mathrm{H}_3^+}^\mathrm{max}$ of H$_3^+$ ions by the product $N_{\mathrm{H}_2} \, N_{\mathrm{H}_2^*} \langle \rho_{\mathrm{H}_2} \rho_{\mathrm{H}_2^*}\rangle_{t,V} D_\gamma$ (see Eq.~\eqref{eq:Nh3}).

\section{Experimental Results}

\begin{figure}
\begin{center}
\includegraphics[width=10 cm]{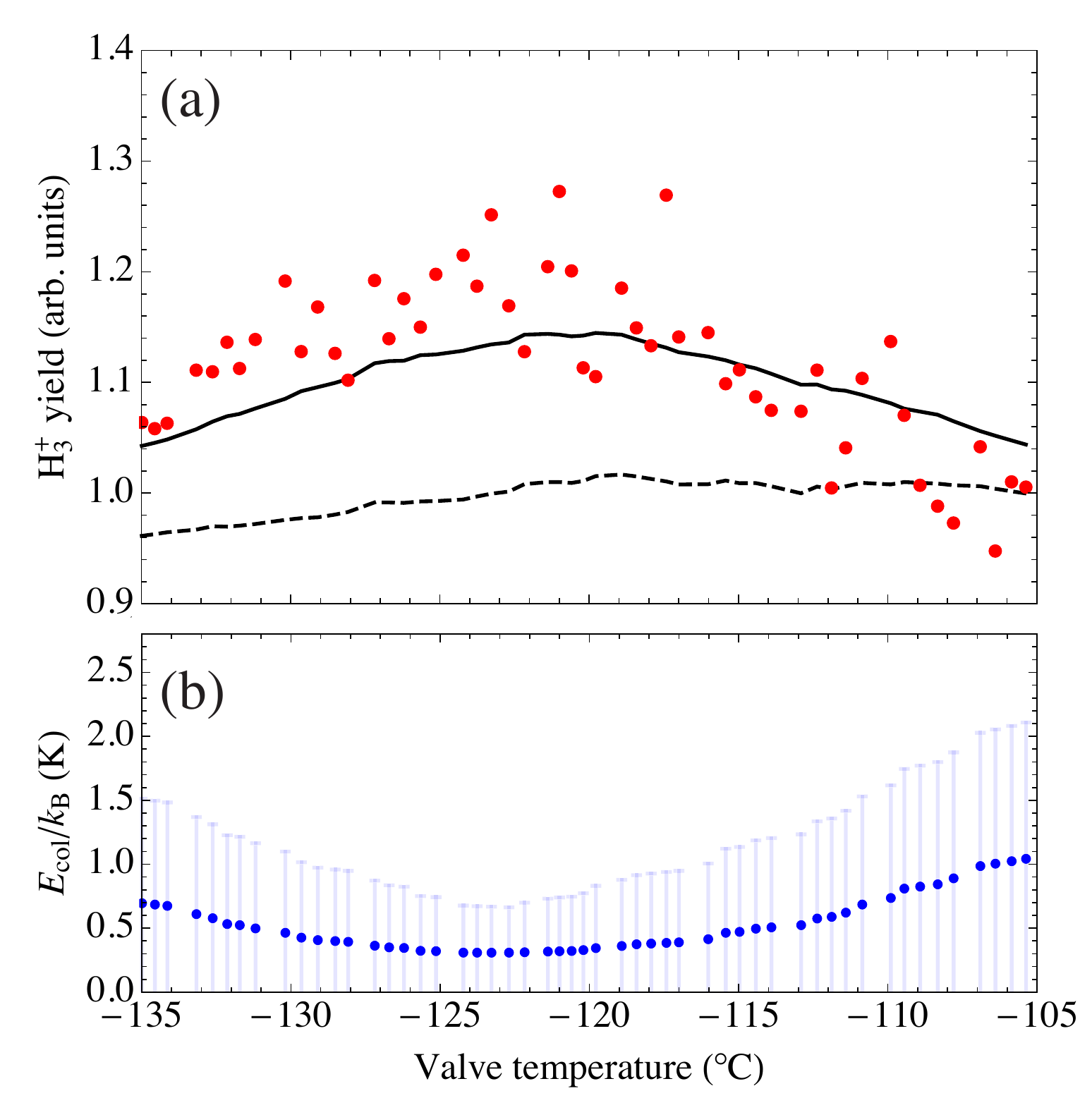}
\caption{\label{fig2} a) Normalized peak number of H$_3^+$ ions $N_\mathrm{H_3^+}^\mathrm{max}$ as a function of the temperature of the ground-state-H$_2$ valve: (red dots) experimental data points for H$_2^*$ Rydberg beam central velocity $v$(H$_2^*$)\,=\,1800~m/s (ion extraction with a two-pulse sequence with $\Delta t = 7 \mu\mathrm{s}$), scaled to 1.0 at -105$^\circ$~C; (dashed line) simulation based on an energy-independent rate coefficient; (full line) simulation based on the rate coefficients predicted by Dashevskaya {\it et al.} and displayed as green line in Fig. 2 of Ref. \cite{dashevskaya16a}. b) Ranges of collision energies (vertical bars) and of mean collision energies (dots) probed experimentally as a function of the temperature of the ground-state-H$_2$ valve when the mean velocity of the Rydberg-H$_2$ beam selected by the RYSDAC is 1800~m/s.}
\end{center}
\end{figure}

In a previous set of experiments \cite{allmendinger16a}, we measured the relative cross section of Reaction~(\ref{eq:reaction}) over the range of collision energies between $k_{\rm B}\cdot 60$~K and $k_{\rm B}\cdot 5$~K, typical for molecular clouds in the interstellar medium, under conditions where the velocity of the Rydberg-H$_2$ beam was always lower than that of the ground-state H$_2$ beam. To study the range of collision energies below $k_{\rm B}\cdot 5$~K in the present study, we slightly accelerated the Rydberg-H$_2$ beam to a velocity of 1800~m/s and varied the collision energies by tuning the temperature of the ground-state H$_2$ valve from $-135\ ^\circ$C, at which point the velocity of the ground state molecules is lower than that of the Rydberg molecules, to $-105\ ^\circ$C, where it is higher. The relative mean velocity of the beams passes through zero at a ground-state-H$_2$-beam valve temperature of $-123\ ^\circ$C. The range of collision energies probed in this experiment is depicted in Fig.~\ref{fig2}b, where the dots and the vertical bars indicate the mean value of the collision energies and the full widths of the distributions of collision energies, respectively, which we know precisely from separate measurements and from the full numerical simulation of the propagation of both beams, as described in detail in Ref.~\cite{allmendinger16a}. When the relative mean velocity of the two beams passes through zero, the mean collision energy is only $k_{\rm B}\cdot 300$~mK, limited by the velocity distribution of the Rydberg H$_2$ beam, as discussed above. We estimate the systematic uncertainty in the determination of the valve temperature at which the relative velocity of the two beams crosses zero to be about 1~K, leading to a negligible uncertainty in the determination of the collision energy (see Fig.~\ref{fig2}b).

Given that the distribution of collision energies is known and that the numbers of reactant molecules ($N_{\mathrm{H}_2}$ and  $N_{\mathrm{H}_2^*}$) and the overlap factor $\langle \rho_{\mathrm{H}_2} \rho_{\mathrm{H}_2^*}\rangle_{t,V,\gamma}$ (see Eq.~(\ref{eq:Nh3})) are independently determined up to a global scaling factor, one can predict the relative yield of H$_3^+$ products for an assumed energy dependence of the reaction rate. If, for instance, one assumes that the reaction rate is independent of the collision energy, as would be the case for a classical Langevin-capture behavior (see Eq.~(\ref{eq:langevin})), the H$_3^+$ product yield depicted as dashed line in Fig.~\ref{fig2}a is obtained. The weak oscillations in this curve have their origin in the small fluctuations of the particle densities and velocities measured experimentally. The fact that this dashed curve significantly  drops below 1 (by about 6\%) at the lowest valve temperatures is the consequence of a reduced detection efficiency of the H$_3^+$ product ions. Because the velocity of the Rydberg-H$_2$ beam is kept constant during the measurement, the mean center-of-mass velocity of the reactants decreases as the ground-state H$_2$ beam velocity is reduced. The H$_3^+$ product ions that are emitted against the propagation direction of the merged beams, and which make the largest part of the H$_3^+$ signal, are detected slightly less efficiently as the center-of-mass velocity is reduced because the fastest H$_3^+$ ions emitted in the backward direction are lost from the detection volume before the product-extraction electric field is applied.

The H$_3^+$-product-ion yields we measure under these conditions are presented as red dots in Fig.~\ref{fig2}a and systematically deviate from the dashed line, the deviation being largest at the valve temperatures for which the collision energy is smallest, i.e. around $-123\ ^\circ$C where the mean collision energy is $k_{\rm B}\cdot 300$~mK, and are smallest at the valve temperatures for which the collision energy is largest, i.e., around $-105\ ^\circ$C where the mean collision energy is $k_{\rm B}\cdot 1.05$~K. At this latter collision energy, the observed reaction rate does not deviate, within the uncertainty of our measurements, from the behavior predicted on  the basis of an energy-independent reaction rate coefficient. The largest deviation between our experimental data and the dashed line is 
about 15\% and unambiguously indicates that the reaction rate coefficient increases as the collision energy decreases towards zero.

\begin{center}
\begin{figure}[tb]
\includegraphics[width=12 cm]{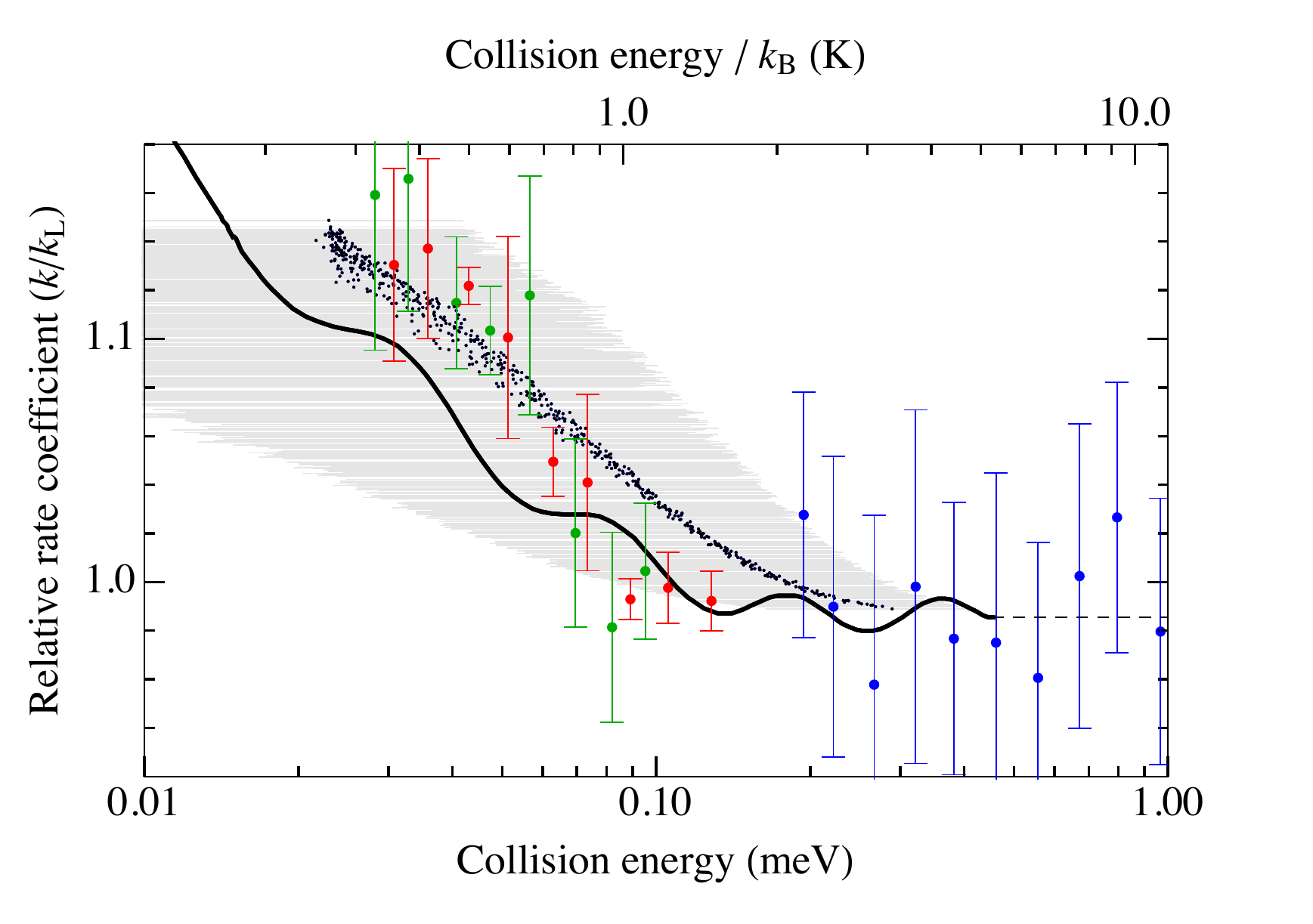}
\caption{Comparison of the energy-dependence of the measured relative rate coefficients $k(E) / k_L$ (color dots) to the calculation by Dashevskaya \textit{et al.} \cite{dashevskaya16a} for normal H$_2$ (75\% H$_2$ in $j=1$ and 25\% H$_2$ in $j=0$) at fixed collision energies (solid line) and for collision energies averaged over the simulated experimental energy distributions (black dots, gray bars indicate one standard deviation). Green dots: two-pulse sequence ($\Delta t = 7 \mu\mathrm{s}$) and H$_2^*$ Rydberg beam central velocity $v$(H$_2^*$)\,=\,1800~m/s. Red dots: single-pulse sequence for $v$(H$_2^*$)\,=\,1700~m/s. Blue dots: single-pulse sequence, $v$(H$_2^*$)\,=\,1540~m/s. The absolute scaling of each experimental data set was chosen to minimize the deviation from the simulation. The simulation (black dots) is based on the experimental parameters of the two-pulse measurement (green dots), but the result is very similar for the other measurements.}
\label{fig3}
\end{figure}
\end{center}

The energy dependence of the ratio $k(E) / (\kappa k_\mathrm{L})$ is extracted from the data by dividing the measured number of H$_3^+$ ions (red dots in Fig.~\ref{fig2}\,a)) by the simulated product-ion yield assuming a constant reaction rate coefficient. The constant $\kappa$ is chosen so that the experimental results match the theoretical results above 1 K (dashed black line in Fig.~\ref{fig2}a)). The experimental data are binned in classes of collision energy. The results of this procedure are depicted by the green dots in Fig.~\ref{fig3}, which show that the rate coefficients rise by about 15\% when the mean collision energy is reduced from $k_{\rm B}\cdot\unit[1.05]{K}$ to $k_{\rm B}\cdot\unit[0.3]{K}$. The vertical bars in Fig.~\ref{fig3} correspond to one standard deviation.

In a second set of measurements, we reduced the mean velocity of the H$_2$ Rydberg molecules to 1700~m/s by adapting the potentials applied to the RYSDAC and repeated the measurement by varying the temperature of the valve used to generate the ground-state H$_2$ beam over the corresponding range. The results of this measurement, analyzed in the same way as described above, are depicted by the red dots in Fig.~\ref{fig3}. Although the experimental conditions of the two sets of measurements are different, the energy dependence of the extracted ratios $k(E) / k_\mathrm{L}$ agrees within the experimental uncertainties, which illustrates the robustness and reliability of our measurement and analysis procedures. The results of a third set of measurements, recorded with a sample of H$_2$ Rydberg molecules prepared at a mean velocity of 1540~m/s and depicted by blue dots in Fig.~\ref{fig3}, does not reveal any systematic energy dependence of the rate coefficients at collision energies higher than $k_{\rm B}\cdot 1$~K and confirms the results previously reported in Ref.~\cite{allmendinger16a}.

In the accompanying article, Dashevskaya  {\it et al.} \cite{dashevskaya16a} report a prediction of the energy-dependent rate coefficients of the reaction ${\rm H}_2^+ + {\rm H}_2\rightarrow {\rm H}_3^+ + {\rm H}$ for a ground-state H$_2$ sample consisting of a mixture of 25\% para-H$_2$ molecules in the $j=0$ rotational level and 75\% ortho-H$_2$ molecules in the $j=1$ rotational level corresponding to the gas sample used in our experiments (see the green curve labeled $\bar{\chi}$ in their figure 2, which is reproduced as full black line in our Fig.~\ref{fig3}). To see whether their data are compatible with our experimental results, we have calculated the H$_3^+$ ion yield from their data using our simulation program and the velocity and density distributions corresponding to our experiments (i.e., using the same input as used to generate the dashed line in Fig.~\ref{fig2} except the reaction rate coefficient). This procedure resulted in the black dots presented in Fig.~\ref{fig3}, where the horizontal gray bars indicate the range of collision energies sampled.  

The simulation indicates that the observed increase of the reaction rate at low collision energy is consistent with the energy dependence of the rate coefficient for the reaction ${\rm H}_2^+ + {\rm H}_2\rightarrow {\rm H}_3^+ + {\rm H}$ calculated by Dashevskaya  {\it et al.} \cite{dashevskaya16a} when averaged over the experimental distribution of collision energies (black dots and gray bars in Fig.~\ref{fig3}). Our experimental results thus reveal for the first time a pronounced and rather sudden departure of the reaction rate of the ${\rm H}_2^+ + {\rm H}_2\rightarrow {\rm H}_3^+ + {\rm H}$ from the behavior predicted on the basis of the classical Langevin-capture model at low-collision energies.
 
\section{Conclusions}

The good agreement between experimental observations and simulations based on the energy-dependent rate coefficients calculated by Dashevskaya  {\it et al.} \cite{dashevskaya16a} leads to the conclusion that the mechanism responsible for the observed enhancement of the rate coefficient for low collision energies has its origin in the interaction between the charge of H$_2^+$ and the rotational quadrupole moment of the ground state of ortho-H$_2$ ($j=1$).  This interaction, which scales with the intermolecular separation $R$ as $1/R^3$, leads to an anisotropic modification of the long-range scattering potential which is dominated by the isotropic charge--induced-dipole coupling (falling of as $1/R^4$). \textit{A priori} all orientations of the rotating H$_2$ molecule (or, equivalently, all values of the projection quantum number $\omega = 0, \pm 1$ of $j$ onto the collision axis) have equal probabilities. At large collision energies, no (re)locking of the ground-state H$_2$ intrinsic angular momentum takes place, and the anisotropic contributions average out. At low collision energies, however, the collision complex can follow the minimum energy trajectory adiabatically, leading to a ``locking'' of the intrinsic rotation of the H$_2$ molecule to the collision axis and an enhanced rate coefficient (see Fig.~\ref{fig3}). In the limit of zero collision energy, which is not probed yet with sufficient precision in our experiments, the rate constant should approach the Bethe-Wigner limit of about 3.6~$k_\mathrm{L}$, as given by $\frac{1}{4}\cdot 2+\frac{3}{4}\cdot 4.18$ for a $\frac{1}{4}$-$\frac{3}{4}$ para-ortho H$_2$ mixture at low temperature~\cite{dashevskaya16a}.

Several aspects of the low-collision-energy behavior predicted theoretically by Dashevskaya  {\it et al.} \cite{dashevskaya16a} remain untested by our experiments, such as the weak oscillations of the reaction rate coefficient at low collision energies, which provide information on the contributions of individual partial waves, and the magnitude of the reaction rate at the lowest energies, which is dominated by the relocking of the angular momentum of the ground-state H$_2$ molecules as the capture rate approaches the s-wave-scattering limit.
We expect that improvements of the signal-to-noise ratio and of the energy resolution of our measurements will make the observation of the predicted oscillations of the rate coefficient possible in the future and will enable us to observe an even larger departure from the classical Langevin-capture model.

\section*{acknowledgments}
We thank Hansj{\"u}rg Schmutz and Josef A. Agner for their help in the development of the experimental setup, and Professor J. Troe and Professor E. Nikitin for making the content of the accompanying article available to us prior to submission. This work is supported financially by the Swiss National Science Foundation under Project Nr.~200020-159848 and by the NCCR QSIT. 

%

\end{document}